\documentclass[aip,jcp,12pt,showkeys,superscriptaddress,showpacs,reprint,raggedbottom,longbibliography]{revtex4-1}
\usepackage{amsmath}
\usepackage{amsthm}
\usepackage{amsfonts}
\usepackage{amssymb}
\usepackage{graphicx}
\usepackage[colorlinks]{hyperref}
\usepackage{dcolumn}
\usepackage{bm}
\allowdisplaybreaks
\usepackage{wrapfig}

\begin{document}
\title
{
Development of composite control-variate stratified sampling approach 
for efficient stochastic calculation of molecular integrals
}
\author{Michael G. Bayne}
\author{Arindam Chakraborty}
\email[corresponding author: ]{archakra@syr.edu}
\affiliation{Department of Chemistry, Syracuse University, Syracuse, New York 13244 USA}
\date{\today} 
%--ABSTRACT--%
\begin{center}
\begin{abstract}
Efficient evaluation of molecular integrals is central 
for quantum chemical calculations. Post Hartree-Fock 
methods that are based on perturbation theory, configuration 
interaction, coupled-cluster, and many-body Green's function 
based methods require access to 2-electron molecular orbital 
(MO) integrals in their implementations. In conventional 
methods, the MO integrals are obtained by the transformation of 
pre-existing atomic orbital (AO) integrals and the computational 
efficiency of AO-to-MO integral transformation has long  been 
recognized as one of the key computational demanding steps in 
many-body methods. In this work, the composite control-variate 
stratified sampling (CCSS) method is presented for calculation 
of MO integrals without transformation of AO integrals. The 
central idea of this approach is to obtain the 2-electron MO 
integrals by direct integration of 2-electron coordinates. 
This method does not require or use pre-computed AO integrals 
and the value of the MOs at any point in space is obtained 
directly from the linear combination of AOs. The integration 
over the electronic coordinates was performed using stratified 
sampling Monte Carlo method. This approach was implemented by 
dividing the integration region into a set of non-overlapping 
segments and performing Monte Carlo calculations on each segment. 
The Monte Carlo sampling points for each segment were optimized 
to minimize the total variance of the sample mean. Additional 
variance reduction of the overall calculations was achieved by 
introducing control-variate in the stratified sampling scheme. 
The composite aspect of the CCSS allows for simultaneous 
computation of multiple MO integrals during the stratified 
sampling evaluation. The main advantage of the CCSS method is 
that unlike rejection sampling Monte Carlo methods such as Metropolis 
algorithm, the stratified sampling  uses all instances of the 
calculated functions for the evaluation of the sample mean. The CCSS 
method is designed to be used for large systems where AO-to-MO 
transformation is computationally prohibitive. Because it is 
based on numerical integration, the CCSS method can be applied 
to a wide variety of integration kernels and does not require 
\textit{a priori} knowledge of analytical integrals.  In this 
work, the developed CCSS method was applied for calculation of 
exciton binding energies in CdSe quantum dots using electron-hole 
explicitly correlated Hartree-Fock (eh-XCHF) method and excitation 
energy calculations using geminal-screened electron-hole interaction 
kernel (GSIK) method. The results from these calculations demonstrate 
that the CCSS method enabled the investigation of excited state 
properties of quantum dots by avoiding the computationally challenging 
AO-to-MO integral transformation step. 
\end{abstract}
\end{center}
%\pacs{31.15.V}
\thispagestyle{plain}

\maketitle

%--------------------------------------------------------------------------------------%
\section{Introduction}
Matrix elements of molecular orbitals (MOs) are central to quantum 
chemical calculations. The MOs form a natural choice for 
single-particle basis functions used in the second-quantized 
representation for many-body post Hatree-Fock (HF) theories.
In the LCAO-MO representation, each 
molecular orbital is represented as a linear combination of 
a set of atomic orbitals. The expansion coefficients of the 
MOs in terms of the AOs are obtained by solving the 
pseudo-eigenvalue Fock equation using the SCF procedure. 
Evaluation of the matrix elements in the MO representation 
requires transformation of the AO integrals. For example, in 
the case of the two-electron Coulomb integral this 
expansion is given as,
\begin{widetext}
    \begin{align}
    \label{eq:2e}
    	[\psi_p(1)\psi_q(1) \vert r_{12}^{-1} \vert \psi_r(2) \psi_s(2)]
    	&=
    	\sum_{\mu\nu\lambda\sigma}
    	C_{\mu p} C_{\nu q} C_{\lambda r} C_{\sigma s}
    	[ \phi_{\mu}(1) \phi_{\nu}(1) \vert r_{12}^{-1} \omega(1,2)  \vert \phi_{\lambda}(2) \phi_{\sigma}(2)] .
    \end{align} 
\end{widetext}
As seen from \autoref{eq:2e}, the transformation formally 
scales as the $4^\mathrm{th}$ power of the number of AO basis 
functions ($N_b$).  
There are various situations where
efficient computation of MO integrals
is required to perform electronic structure calculations. 
For example, application of many-body theories
such as configuration interaction (CI),\cite{Fales20154708,levine_2017_direct} 
many-body perturbation theory (MBPT),\cite{Katouda20162623,Ohnishi2015333} 
and couple-cluster theory (CC)~\cite{kowalski_2017_highly} for large chemical 
systems need fast and efficient access to these MO integrals.
\par
Efficient calculation on MO integrals is a 
recurrent theme in increasing the 
efficiency of the electronic structure calculations. 
The transformation can be accelerated 
by performing it in parallel and various parallelization 
algorithms have been developed.~\cite{lopes_2017_fast,piccardo_2017_full,wirz_2017_resolution}
The computational cost can also be 
reduced using rank-reduction techniques
such as
resolution-of-identity\cite{bostrom_2009_ab,sabin_1979_some,
whitten_1973_coulombic,lee_1994_coupled,
headgordon_2006_linear,
knowles_2003_fast,
komornicki_1993_use,
feyereisen_1993_integral,
weigend_2002_fully,takeshita_2017_stochastic}  
and 
Cholesky decomposition.\cite{johansen_2008_cholesky,lindh_2007_low,
linderberg_1977_simplifications,pedersen_2003_reduced,
roeggen_1986_on}
In a series of papers, Martinez \textit{et al.} 
have developed 
the tensor-hypercontraction  
approach\cite{martinez_2017_analytical,
martinez_2017_atomic,martinez_2016_atomic,
martinez_2012_tensor,sherrill_2012_tensor,
sherrill_2012_tensorII,martinez_2012_comm,
martinez_2013_quartic,sherrill_2013_discrete,
martinez_2013_exact,
martinez_2013_tensor,
martinez_2014_comm,
martinez_2015_tensor} 
that has enabled significant 
reduction in the computational cost of 
electron-repulsion integrals (ERI). 
A current review of the various ERI
techniques has been presented by 
Peng and Kowalski.\cite{kowalski_2017_highly} 
\par 
Efficient evaluations of MO integrals are also required in 
explicitly correlated methods\cite{valeev_2006_r12, 
varganov2010variational,
Cagg_2014_strongly,Jeszenszki_2015_local,
Rassolov19993672,Rassolov2009,
Rassolov2011,nichols_2013_description} 
where the evaluation of the 
r12-kernel in AO representation 
is not readily available or is not computationally efficient. 
For a n-body operator, the AO-to-MO transformation scales as $N_b^{2n}$
and becomes computationally expensive for n-body operators
when $n>2$ because of steep scaling with
respect to the number of AO basis functions.
This has found to be especially true for 
explicitly correlated methods for treating
electron-electron,\cite{valeev_2006_r12,varganov_2010_variational,
tenno_2017_perspective,hirata2014second,
hirata_2015_general,johnson2016monte,
willow_2014_brueckner,
willow2014stochastic} 
electron-proton,\cite{hs_2017_development,
hs_2015_dependence,hs_2015_quantum,
hs_2012_multicomponent,hs_2011_diabatization,
hs_2011_derivation,hs_2011_alternative,
hs_2008_density,hs_2008_development,
hs_2008_inclusion,hs_2006_explicitly,
hs_2005_alternative,hs_2004_electron} 
and  electron-hole\cite{elward_2012_calculation,
elward_2013_effect,chakraborty_2016_shape,
ellis_2016_development,bayne_2016_construction,
elward_2015_effect,bayne_2014_infinite,
elward_2014_optical,blanton_2013_development,
elward_2012_variational,elward_2012_investigation}  
many-body theories.  
One approach to avoid the transformation of the AO integrals is to  
use real-space representation and to evaluate the
MO integrals numerically. This procedure requires
evaluation of the MOs at any position in the 3D space
which can be evaluated from the AO expansion,
\begin{align}
	\psi_p(\mathbf{r}) 
	&=
	\sum_{\mu}
	C_{\mu p} \phi_{\mu}(\mathbf{r}) .
\end{align}
This strategy has been used very successfully in 
quantum Monte Carlo methods
\cite{filippi_1996_multiconfiguration,filippi_2013_ground,
filippi_2014_practical,filippi_2004_excitations,filippi_2016_multiple, 
ceperley_2007_quantum, lester2002recent,nightingale1998quantum} 
where 
evaluation of individual MO integrals can be completely 
avoided and the entire many-electron integral is evaluated 
directly in real-space representation using
Markov chain Monte Carlo (MCMC) implementation. 
The MCMC implementation was also shown to be used 
in the context of perturbation theory 
in a series of articles by Hirata \textit{et al.} 
\cite{hirata2014second,hirata_2015_general,
johnson2016monte,
willow_2014_brueckner,willow2014stochastic} 
in which MCMC techniques were used for the 
evaluation of MP2-F12 energies. 
\par
In this work we present the composite control-variate stratified sampling (CCSS)
Monte Carlo method for efficient calculation of MO integrals. 
The accuracy of stochastic evaluation of integrals can be
systematically  improved
by reducing the variance of the calculation. 
In the CCSS method, 
we have combined both control-variate and stratified sampling strategies
for variance reduction. The CCSS method was used in conjunction with the 
electron-hole explicitly correlated Hartree-Fock method (eh-XCHF)
for the calculation of exciton binding energies and excitation energies
in CdSe clusters and quantum dots.  
%--------------------------------------------------------------------------------------%
\section{Theory}
\label{sec:theory}

\subsection{Coordinate transformations}
We start by defining the following general two-electron integral 
of the following form,
\begin{align}
\label{eq:int1}
	I_{pqrs}
	&=	
	\iint\limits_{-\infty}^{+\infty} d\mathbf{r}_1 d\mathbf{r}_2 \,
	\Lambda_{pq}(1) \Lambda_{rs}(2) r_{12}^{-1} \omega(1,2) ,
\end{align} 
where $\Lambda_{pq} = \psi_p \psi_q$ and $\Lambda_{rs} = \psi_r \psi_s$.
We will transform the two-electron 
coodinate system into intracular and 
extracular coordinates,
\begin{align}
	\mathbf{r}_{12}
	&= 
	\mathbf{r}_{1}
	- \mathbf{r}_{2} \\
	\mathbf{R}
	&=
	\frac{1}{2} 
	(\mathbf{r}_{1} + \mathbf{r}_{2}) .
\end{align}
%The Jacobian for this transformation is one
The Jacobian for this transformation is,
\begin{align}
	d\mathbf{r}_1 	d\mathbf{r}_2
	&=  d\mathbf{R} 	d\mathbf{r}_{12} .
\end{align}
In the next step, we will transform into 
spherical polar coordinates,
\begin{align}
\label{eq:polar1}
	d\mathbf{r}_{12}
	&=
	r_{12}^2 \sin(\theta_{12}) dr_{12} d\theta_{12} d\phi_{12} \\
	d\mathbf{R}
	&=
	R^2 \sin(\Theta) dR d\Theta d\Phi .
\end{align}
Using \autoref{eq:polar1}, the integral \autoref{eq:int1} 
is, 
\begin{align}
\label{eq:int2}
	I_{pqrs}
	=	
	&\iint\limits_{0}^{\infty} dR dr_{12} r_{12} R^2
	\iint\limits_{0}^{\pi} d\Theta d\theta_{12} \sin^2\Theta \sin^2\theta_{12} \notag\\
	&\iint\limits_{0}^{2\pi} d\Phi d\phi
	\Lambda_{pq}(1) \Lambda_{rs}(2) \omega(1,2) .
\end{align}
The transformation to the spherical polar coordinates allows us to 
analytically remove the $r_{12}^{-1}$ singularity in the integration kernel. 
In many applications, the operator $\omega(1,2)$ might depend 
only on $r_{12}$ in which case it can be moved out of the integration over 
the angular coordinates. 
For performing Monte Carlo calculation to evaluate this integral 
numerically, it is convenient to transform the integration limits
to $[0,1]$. 
Now we will perform a third coordinate transformation 
and transform the integration domain to $[0,1]$ limits. 
This is done mainly to aid in the numerical evaluation 
of the integral using Monte Carlo techniques.
We define a new set of coordinates $(\mathbf{t}=\{t_1,t_2,\dots,t_6\})$
where each coordinate is in the range $t \in [0,1]$. 
The radial and angular coordinates are transformed as,
\begin{align}
\label{eq:t1}
	r           &= \frac{t}{1-t} \\
	\theta   &= \frac{t}{\pi} \\
	\phi       &=  \frac{t}{2\pi} .
\end{align}
The associated Jacobians are,
\begin{align}
\label{eq:t2}
	dr &= \frac{1}{(1-t)^2} dt \\
	d\theta   &= \frac{1}{\pi} dt \\
	d\phi       &=  \frac{1}{2\pi} dt .
\end{align}
In the $t$-space, the 
expression for $I_{pqrs}$ can be expressed compactly as, 
\begin{align}
	I_{pqrs}
	&=
	\int_{0}^{1} d\mathbf{t} f(\mathbf{t}) .
\end{align}
The integral kernel  $f(\mathbf{t})$
is obtained by substituting \autoref{eq:t1} and \autoref{eq:t2} into \autoref{eq:int2},
\begin{widetext}
    \begin{align}
    	f(\mathbf{t})
    	&=
    	\left(\frac{1}{2\pi^2} \right)^2
    	\frac{t_1}{(1-t_1)^3} \frac{t_2^2}{(1-t_2)^4}
    	\sin(t_3/\pi) \sin(t_4/\pi)
    	\Lambda_{pq}(\mathbf{t}) \omega(\mathbf{t}) \Lambda_{rs}(\mathbf{t}) ,
    \end{align}
\end{widetext}
where $t_1$ and $t_2$ corresponds to $r_{12}$ and $R$, respectively, and the 
remaining $t_i$ are angular coordinates. 
Using Monte Carlo, the estimation of $I_{pqrs}$ is then given by the following expression,
\begin{align}
\label{eq:int3}
	I_{pqrs} 
	\approx
	\mathbb{E}[f]  \pm \sqrt{\frac{\mathbb{V}[f]}{N_\mathrm{S}}} ,
\end{align}
where $N_\mathrm{S}$ is the number of sampling points and $\mathbb{E}$  is the expectation value. $\mathbb{V}$ is the variance defined by \autoref{eq:eval1}
and \autoref{eq:vval1}, respectively, and is shown below, 
\begin{align}
\label{eq:eval1}
	\mathbb{E}[f]
	&=
	\frac{1}{N_\mathrm{S}}
	\sum_{i=1}^{N_\mathrm{S}}
	f(\mathbf{t}_i)
\end{align} 
\begin{align}
\label{eq:vval1}
	\mathbb{V}[f]
	&=
	\mathbb{E}[f^2] -\mathbb{E}[f]^2 .
\end{align}
A summary of key relationships 
between expectation value and variance
that is relevant to this derivation 
is provided in appendix A.
As seen from \autoref{eq:int3}, the error in the 
numerical estimation of the integral depends on the
variance, hence it is desirable to reduce the overall 
variance of the sampling to obtain an accurate value of the 
integral. 
In this work, we have combined stratified sampling approach 
with the control-variate method to achieve variance reduction.

\subsection{Stratified sampling}
Stratified sampling is a successful strategy to 
reduce the variance of the overall estimate of the calculation.
This is a well-know technique that has been described
earlier in previous publications.\cite{Rubinstein2016,pruneau2017data,Kroese2013,Fishman2013,Kalos2008}
Here, only the key
features of the method that are directly related to 
this work are summarized below. 
Stratified sampling can be implemented using 
both constant-volume or different-volume
segments, and in this work we have used only 
the constant volume version. 
In the constant-volume approach, the integration domain $\Omega$ 
of the integration region is uniformly divided among 
non-overlapping segments (\autoref{eq:seg1}), 
\begin{align}
\label{eq:seg1}
	\Omega = \sum_{\alpha = 1}^{N_\mathrm{seg}} \Omega_\alpha .
\end{align}
We have used a direct-product 
approach for generation of the segments. 
Along each t-dimension, the region $[0,1]$
was divided equally into $2^m$ segments.
The segments for the 6-dimension was obtained by the 
direct-products of the 1-dimensional segments. 
This procedure resulted in a total of $N_\mathrm{seg} = 2^{6m}$ 
number of 6D segments.  
The sample mean and variance associated with each 
segment $\alpha$ is given as,
\begin{align}
\label{eq:mu1}
	\mu_\alpha
	=
	\mathbb{E}[f_\alpha] 
	=
	\frac{1}{N_\mathrm{S}^\alpha} 
	\sum_{\mathbf{t} \in \Omega_\alpha}
	f(\mathbf{t}) ,
\end{align} 
where $N_\mathrm{S}^\alpha$
is the number of sampling points used in the evaluation 
of the expectation value for segment $\alpha$. 
The notation $\mathbf{t} \in \Omega_\alpha$
implies that points only in the domain $\Omega_\alpha$
should be used for evaluation of the expectation value $\mathbb{E}$.
Analogous to \autoref{eq:vval1}, the variance associated with each segment is defined as,
\begin{align}
\label{eq:sig1}
	\sigma_\alpha^2
	=
	\mathbb{V}[f_\alpha] 
	= 
	\mathbb{E}[f^2_\alpha] - \mathbb{E}[f_\alpha]^2 .
\end{align}
The estimate of the total expectation value is obtained by the 
average over all the segments.
Mathematically, this can be expressed as,
\begin{align}
\label{eq:mu2}
	\mathbb{E}[f]
	= \mu 
	=
	\frac{1}{N_\mathrm{seg}} \sum_{\alpha=1}^{N_\mathrm{seg}}
	\mu_\alpha .
\end{align}
In \autoref{eq:mu2}, 
partial averages from all the segments contribute equally
because all the segments have exactly identical volumes.
For cases where segments have different volumes, the above expression should be replaced by a weighted average. 
To calculate the variance on $\mu$ we will 
use the relationship that the variance of 
sum of two random variates are related to each 
other by their covariance (derived in \autoref{eq:var3}) as shown below,
\begin{align}
\label{eq:vval2}
	\mathbb{V}[ \sum_i a_i X_i]
	&= 
	\sum_{ij} a_i a_j \mathbb{C}[X_i,X_j] ,
\end{align}
where covariance $\mathbb{C}$  
defined as, 
\begin{align}
\label{eq:covar2}
	\mathbb{C}[X,Y]
	&=
	\mathbb{E}[XY]-\mathbb{E}[X]\mathbb{E}[Y] .	
\end{align}
Using the relationship in \autoref{eq:covar2}, 
\begin{align}
	\mathbb{V}[\mu] 
	&= 
	\mathbb{V}[\frac{1}{N_\mathrm{seg}} \sum_{\alpha=1}^{N_\mathrm{seg}}
	\mu_\alpha] \\
	&= 
	\frac{1}{N_\mathrm{seg}^2}
	\sum_{\alpha \beta}^{N_\mathrm{seg}}
	\mathbb{C}[\mu_\alpha,\mu_\beta] .
\end{align}
Because the sampling of any two segments are 
completely uncorrelated, 
all the off-diagonal elements of the covariance matrix 
will be zero,
\begin{align}
\label{eq:covar3}
	\mathbb{C}[\mu_\alpha,\mu_\beta] 
	&= \mathbb{V}[\mu_\alpha] \delta_{\alpha \beta} .
\end{align}
Using \autoref{eq:covar3} and result from \autoref{eq:var4},
\begin{align}
\label{eq:covar4}
	\mathbb{V}[\mu] 
	&= 
	\frac{1}{N_\mathrm{seg}^2}
	\sum_{\alpha}^{N_\mathrm{seg}}
	\mathbb{V}[\mu_\alpha] .
\end{align}
The result from \autoref{eq:covar4} implies that the 
variance of the mean always decreases with increasing 
number of segments. 
The variance of the segment mean, $\mu_\alpha$, 
is related to related to sample variance by the 
following relationship (derived in \autoref{eq:var5}),\cite{Rubinstein2016,pruneau2017data,Kroese2013,Fishman2013,Kalos2008}
\begin{align}
	\mathbb{V}[\mu_\alpha] 
	&=
	\frac{\mathbb{V}[f_\alpha]}{N_\mathrm{S}^\alpha} .
\end{align}
This implies,
\begin{align}
\label{eq:var1}
	\mathbb{V}[\mu] 
	&= 
	\frac{1}{N_\mathrm{seg}^2}
	\sum_{\alpha }^{N_\mathrm{seg}}
	\frac{\mathbb{V}[f_\alpha]}{N_\mathrm{S}^\alpha} .
\end{align}
The central idea of stratified sampling is to optimize the
distribution of sampling points across all segments to 
reduce the variance in the mean. 
To achieve this, a normalized weight factor, $w_\alpha$,
is associated with each segment
and is given by,
\begin{align}
	\sum_\alpha^{N_\mathrm{seg}} w_\alpha = 1
	\quad \textrm{and} \quad w_\alpha \ge 0 .
\end{align}
The number of sampling points for each segment is 
given by a fraction of the total number of sampling points, 
\begin{align}
	N_\mathrm{S}^\alpha 
	&= 
	w_\alpha N_T .
\end{align} 
Substituting in \autoref{eq:var1},
\begin{align}
\label{eq:var2}
	\mathbb{V}[\mu] 
	&= 
	\frac{1}{N_\mathrm{seg}^2 N_\mathrm{T}}
	\sum_{\alpha}^{N_\mathrm{seg}}
	\frac{1}{w_\alpha} \mathbb{V}[f_\alpha] .
\end{align} 
It can be shown that the optimal distribution of points is achieved 
by selecting the weights proportional to the standard-deviations 
of each segment,\cite{Rubinstein2016,pruneau2017data,Kroese2013,Fishman2013,Kalos2008}
\begin{align}
\label{eq:wt1}
	\min_{\mathbf{w}} \mathbb{V}[\mu] 
	\rightarrow 
	w^\mathrm{opt}_\alpha
	= \frac{ \sqrt{\mathbb{V}[f_\alpha]}   }{\sum_\beta^{N_\mathrm{seg}} \sqrt{\mathbb{V}[f_\beta]} } .
\end{align}
The above equation very nicely illustrates the
intuitive logic behind stratified sampling that
segments with higher variance (or standard deviation)
should receive proportionally more sampling points 
than regions with lower variance. 
The optimized distribution of weights and 
inverse dependence on the number of segments are
the two main reasons why stratified sampling 
is an effective technique for variance reduction.  

\subsection{Variance reducing using control-variate}
Control-variate is another strategy 
that has been used in past for reducing the variance of 
Monte Carlo calculations.\cite{Rubinstein2016,pruneau2017data,Kroese2013,Fishman2013,Kalos2008} 
In this work, we  have
incorporated control-variate technique in our 
stratified sampling calculations. 
In control-variate methods, we start with a function 
(denoted as $f_0(\mathbf{t})$) whose
integral is known in advance,
\begin{align}
	I_{pqrs}^0
	&=
	\int\limits_{0}^{1} d\mathbf{t} f_0(\mathbf{t}) .
\end{align} 
We then add and subtract this quantity from the integral to be evaluated, 
\begin{align}
	I_{pqrs}
	&= 
	\int\limits_{0}^{1} d\mathbf{t} f(\mathbf{t})
	+ 
	\eta
	\left[
	I_{pqrs}^0 - \int\limits_{0}^{1} d\mathbf{t} f_0(\mathbf{t})
	 \right] ,
\end{align}
where $\eta$ is a yet to be determined scaling parameter.
Rearranging we get,
\begin{align}
	I_{pqrs}
	&= 
	\eta 	I_{pqrs}^0
	+
	\int\limits_{0}^{1} d\mathbf{t}
	\left[ f(\mathbf{t}) - \eta f_0(\mathbf{t}) \right] .
\end{align}
The optimum value of the 
scaling parameter $\eta$ is obtained 
by minimizing the variance given in \autoref{eq:var2},
\begin{align}
\label{eq:eta1}
	\min_{\eta} \mathbb{V}[\mu] 
	\rightarrow \eta_\mathrm{opt} .
\end{align}
Because of the above minimization, the variance 
obtained from control-variate sampling is always lower
or equal to the variance obtained without using 
control-variate,
\begin{align}
	\big( \mathbb{V}[\mu] \big)_{\eta_\mathrm{opt}}
	\le 
	\big( \mathbb{V}[\mu] \big)_{\eta = 0} .
\end{align}
\par
Conceptually, control-variate method allows us to 
perform Monte Carlo calculation only on the 
component of the $f$ that is different from $f_0$.
For integration over molecular integrals,
one of the simplest control-variate function is the 
overlap integral, 
\begin{align}
	f_0(1,2)
	&= 
	\left[\chi_p(1) \chi_q(1) \right]
	\left[\chi_r(2) \chi_s(2) \right] \\
	I_{pqrs}^0 
	&= \delta_{pq} \delta_{rs} .
\end{align}
In the case that the underlying AO integrals are available,
a better estimate of $f_0$ can be constructed. 
For example, collecting only the diagonal elements of 
the $\sum_{\mu \nu \lambda \sigma}$ in \autoref{eq:2e},
the control-variate function $f_0$ can be defined as,
\begin{align}
	f_0(1,2)
	&=
	\sum_{\mu}^{N_\mathrm{b}}
	C_{\mu p} C_{\mu q} C_{\mu r} C_{\mu s}
	 \phi_{\mu}(1) \phi_{\mu}(1)  r_{12}^{-1}  \phi_{\mu}(2) \phi_{\mu}(2) .
\end{align}
The value of the integral $I_0$ is obtained analytically from the underlying AO integrals, 
\begin{align}
	I_{pqrs}^0
	&=
	\sum_{\mu}^{N_\mathrm{b}}
	C_{\mu p} C_{\mu q} C_{\mu r} C_{\mu s}
	 [\phi_{\mu}(1) \phi_{\mu}(1)  \vert  r_{12}^{-1} \vert  \phi_{\mu}(2) \phi_{\mu}(2)] .
\end{align}
We note that unlike $I_\mathrm{pqrs}$, evaluation of $I_{pqrs}^0$
is linear in terms of number of AO basis function $N_\mathrm{b}$.

\subsection{Composite control-variate stratified sampling}
In most applications,
matrix elements of a set of molecular orbitals are needed 
for performing electronic structure calculations.
Although in principle the control-variate stratified sampling 
method presented above can be applied 
for evaluation of each matrix element, however, 
such an approach is computationally inefficient. 
A more efficient approach is to 
evaluate the integrals simultaneously for all the 
matrix elements. We call this approach the 
composite control-variate stratified sampling (CCSS) 
and is described as follows. 
\par
We start with set of MO indicies for which the
integrals are needed to be evaluated,
\begin{align}
	\mathcal{Z}
	&= 
	\{ (p_1q_1r_1s_1),(p_2q_2r_2s_2),\dots,   \} .
\end{align}
If all the MO integrals are needed, this set will be a 
set of all symmetry unique indicies. 
All index combination 
from $\mathcal{Z}$ which are 
known to be zero because of
symmetry arguments are also eliminated 
from the set. 
We will use the 
collective index $K$ to enumerate the individual 
elements of set $\mathcal{Z}$,
\begin{align}
	\mathcal{Z}
	&= 	\{ z_K \} .	
\end{align} 
Because the domain of the integration is 
identical for all the indicies,
all the integrals can be evaluated simultaneously,
\begin{align}
	I_{K}
	&= 
	\eta^K 	I_{K}^0
	+
	\int\limits_{0}^{1} d\mathbf{t}
	\left[ f^K(\mathbf{t}) - \eta^K f_0^K(\mathbf{t}) \right] .
\end{align}
In terms of segments,
\begin{align}
	I_{K}
	&= 
	\eta^K 	I_{K}^0
	+
	\frac{1}{N_\mathrm{seg}}
	\sum_{\alpha}^{N_\mathrm{seg}}
	\mathbb{E}\left[ f^K - \eta^K f_0^K\right] .
\end{align}
The expectation value for each segment
will be evaluated using $N_\mathrm{S}^\alpha$
number of sample points whose distribution is 
defined using the weights obtained in \autoref{eq:wt1}.
However, because each segment is now associated with 
$N_K$ number of functions, there are $w^K$ weights associated with 
each segment. 
In the CCSS method, we renormalize the weights
by choosing the maximum weight associated with 
all the functions for a given segment.
Mathematically, this is described by the 
following equations, 
\begin{align}
\label{eq:wt2}
	x_\alpha^\mathrm{opt}
	&=
	\max_K
	\{ w_{\alpha,K}^\mathrm{opt} \} \\
	w_\alpha^\mathrm{opt}
	&=
	\frac{x_\alpha^\mathrm{opt}}{\sum_\beta x_\beta^\mathrm{opt}} .	
\end{align}

\subsection{Precomputation, run-time computation, and parallelization }
In the CCSS method, because the same set of 
molecular orbitals will be used for calculations 
of all the integrals in set $\mathcal{Z}$, it
is computationally efficient to compute them 
once and use them for all functional evaluations. 
In a single Monte Carlo step in a given segment, 
first a random vector $\mathbf{t} \in \Omega_\alpha$
is obtained and all the MOs at $\mathbf{t}$ are evaluated 
and stored in a vector $\mathbf{v}$ of size $N_\mathrm{MO}$.
The functions $f^K$ and $f^K_0$ are then 
built by reading values from vector $\mathbf{v}$. 
These simple steps result in significant savings in
computation time because it avoids repeated evaluations of 
MO values at point $\mathbf{t}$ for each function evaluation in set $\mathcal{Z}$.  
\par
The implementation of the CCSS method requires the determination 
of two run-time parameters $\eta^K$ and  $w_\alpha^\mathrm{opt}$ 
defined in \autoref{eq:eta1} and  \autoref{eq:wt2}, respectively. 
Instead of evaluating them for the entire run, these parameters
were determined using data 
from the first 10\% of the run and were kept fixed for the remaining 
90\% of the calculation.
As seen from \autoref{eq:wt2}, the evaluation of the weights for 
each segment requires information from all the segments. 
By making these weights constant for the 90\% of the run time 
allows for efficient parallization of the CCSS method by 
completely decoupling information exchange among the segments. 
Consequently, this enables Monte Carlo steps for each segment
to be performed in parallel. This strategy was found to significantly reduce the computational time of the overall calculation.
%--------------------------------------------------------------------------------------%
\section{Results}
\label{sec:results}
\subsection{Electron-hole interaction in CdSe quantum dots with dielectric screening}
\begin{figure*}[!ht]
  \begin{center}
      % left bottom right top
      \includegraphics[trim=0cm 0cm 0cm 0cm,scale=0.7]{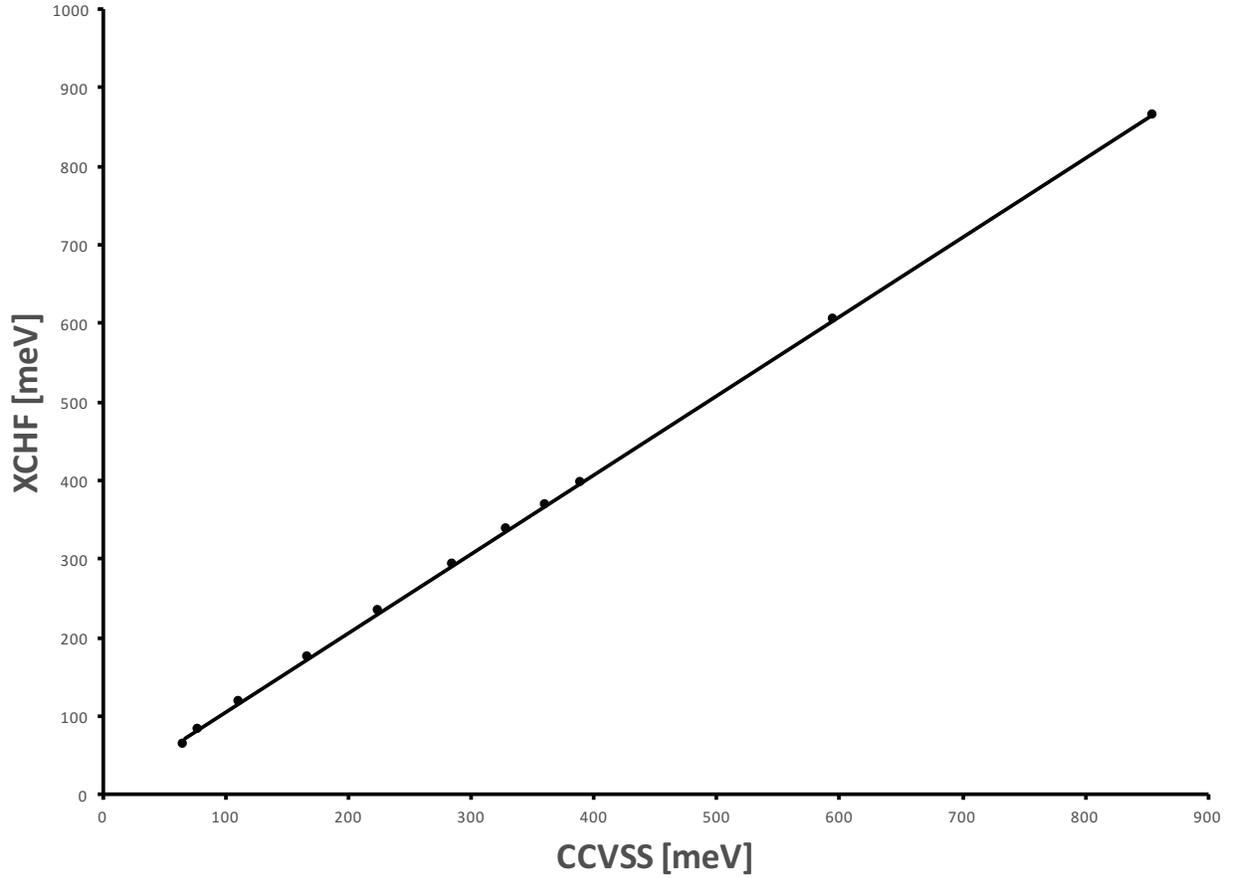}  \\
      \caption{\textbf{Binding energies in meV of CdSe quantum dots ranging in size from 
                       1 nm to 20 nm in diameter of the XCHF method on the y-axis and this 
                       work on the x-axis. The trendline in this graph has a slope of 1.0072}.}
    \label{fig:xchfccvss}
  \end{center}
\end{figure*}

\begin{figure*}[!ht]
  \begin{center}
      % left bottom right top
      \includegraphics[trim=0cm 0cm 0cm 0cm,scale=0.7]{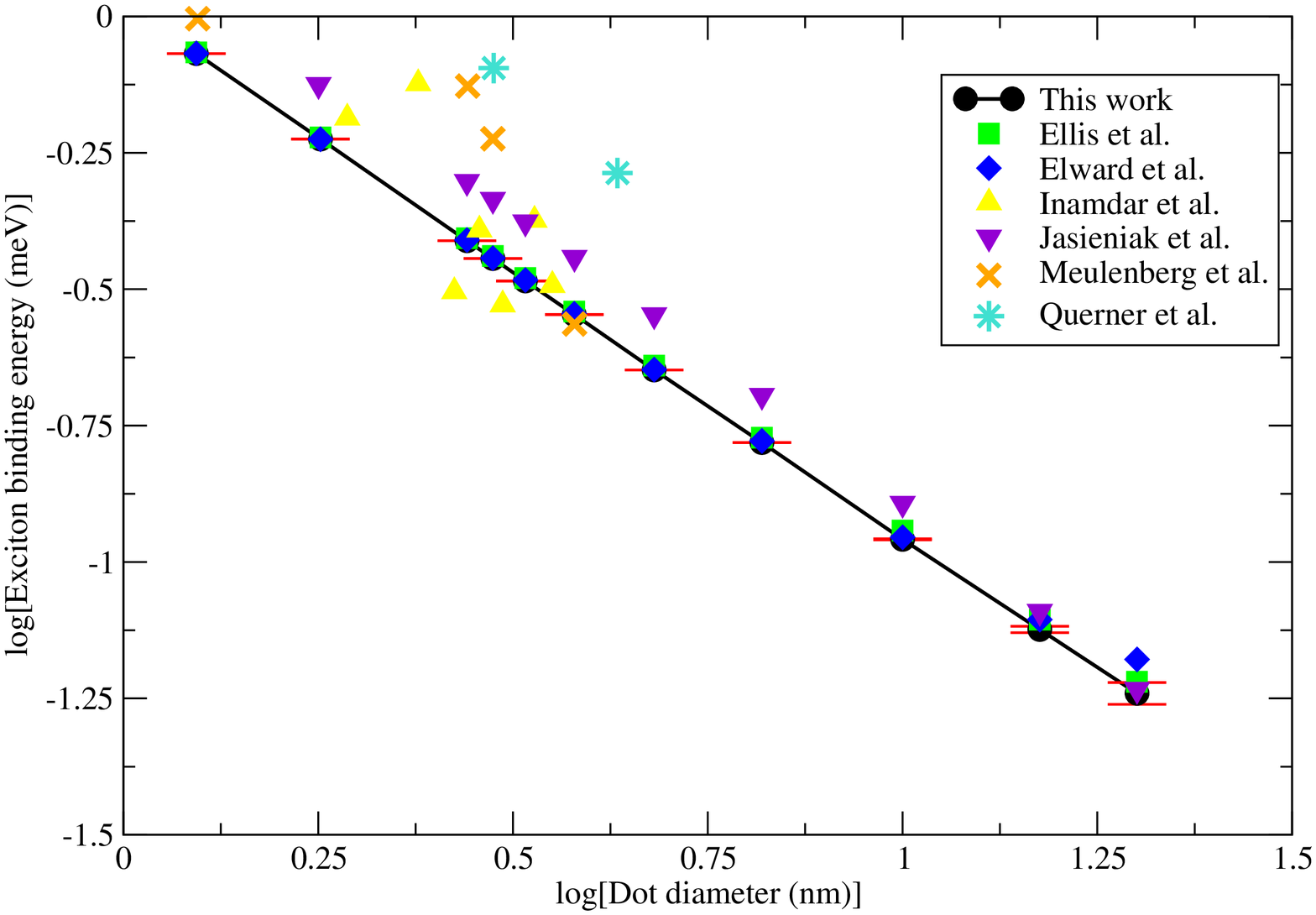}  \\
      \caption{\textbf{Binding energies in meV of CdSe quantum dots ranging in size from 
                       1 nm to 20 nm in diameter of this work compared with Ellis et al.,\cite{ellis_2017_investigation} 
                       Elward et al.,\cite{elward_2013_effect} Inamdar et al.,\cite{inamdar_2008_determination} 
                       Jasieniak et al.,\cite{jasieniak_2011_size}  
                       Muelenberg et al.,\cite{meulenberg_2009_determination} and 
                       Querner et al.\cite{querner_2005_size} 
                       For the CCSS method, red error bars are shown for the exciton binding energy 
                       calculations. 
                       }}
    \label{fig:ccvss_compare}
  \end{center}
\end{figure*}

The CCSS method was used for calculating the exciton binding energies in
a series of CdSe quantum dots using the electron-hole explicitly-correlated 
Hartree-Fock (eh-XCHF) method. 
The eh-XCHF method has been successfully used before \cite{elward_2013_effect} for  investigation of excitonic 
interactions in QDs and only a brief summary relevant to the 
CCSS method is presented here.  
In the eh-XCHF method, the electronic excitation in the QD is described 
using the quasiparticle representation. The electron-hole integration 
is represented using the following effective quasiparticle Hamiltonian, 
\begin{align}
\hat H_\mathrm{eh} &= \sum_{ij} \langle i|\frac{-{\hbar}^2}{2m_{\mathrm{e}}} \nabla^2+v^\mathrm{e}_{\mathrm{ext}}|j\rangle e^\dagger_i e_j \\
& + \sum_{ij} \langle i|\frac{-{\hbar}^2}{2m_{\mathrm{h}}} \nabla^2+v^\mathrm{h}_{\mathrm{ext}}|j\rangle h^\dagger_i h_j \\
& + \sum_{iji^\prime j^\prime} K^\mathrm{eh}_{iji'j'} 
e^\dagger_i e_j h^\dagger_{i^\prime} h_{j^\prime} \\
& + \sum_{ijkl} w^\mathrm{ee}_{ijkl} e^\dagger_i e^\dagger_j{e}_l{e}_k 
+ \sum_{ijkl} w^\mathrm{hh}_{ijkl} h^\dagger_i h^\dagger_j{h}_l{h}_k ,
\end{align}
where the unprimed and primed indicies represent
quasielectron and quasihole states, respectively. 
The attractive electron-hole interaction, $K^\mathrm{eh}$, is the
principle component that results in exciton binding and in these 
calculations, $K^\mathrm{eh}$ was approximated using 
static dielectric screening developed by Wang and Zunger for CdSe QDs.\cite{wang_1996_pseudopotential}
The electron-hole wave function was represented using the 
eh-XCHF ansatz which is defined as,  
\begin{align}
\Psi_{\mathrm{eh-XCHF}} = \hat G \Phi^{\mathrm{e}} \Phi^{\mathrm{h}},
\end{align}
where,
\begin{align}
\hat G = \sum_{i=1}^{N_e} \sum_{j=1}^{N_h} g(i,j),
\end{align}
and $g$ is a linear combination of Gaussian-type geminal functions,
\begin{align}
\label{eq:g1}
	g(1,2) = \sum_{k=1}^{N_\mathrm{g}} b_k e^{-\gamma_k r_{12}^2} .
\end{align}
In the eh-XCHF method the function $g$ is obtained by the following minimization procedure, 
\begin{align}
\mathrm{E} = 
\min_{g} \dfrac{\langle \Phi^{\mathrm{e}}\Phi^{\mathrm{h}}\vert{\hat G^\dagger}{\hat H_\mathrm{eh}}{\hat G}\vert\Phi^{\mathrm{e}}\Phi^{\mathrm{h}}\rangle}{\langle \Phi^{\mathrm{e}}\Phi^{\mathrm{h}}\vert\hat G^\dagger  \hat G\vert\Phi^{\mathrm{e}}\Phi^{\mathrm{h}}\rangle}.
\end{align}
The exciton binding energy is calculated as the difference between the 
interaction and non-interacting energies,
\begin{align}
E_{\mathrm{EB}} = \langle E_{\mathrm{non-interacting}} \rangle - \langle E_{\mathrm{exciton}} \rangle .
\end{align}
The eh-XCHF formulation requires
matrix elements of molecular orbitals involving 
the Coulomb operator $r_{12}^{-1}$ and the Gaussian-type geminal function $g$
and is an ideal candidate to test the CCSS method. 
In the previous applications of the eh-XCHF method,\cite{elward_2013_effect} 
these integrals were
evaluated using analytical geminal integrals.  
For testing the CCSS implementation, we calculated the exciton binding energies
in CdSe clusters and compared with the previously
reported\cite{elward_2013_effect}  exciton binding energies obtained using analytical AO integrals. 
The results from the CCSS methods are summarized in \autoref{fig:xchfccvss}.
The results show that the exciton binding energies obtained using the CCSS
method are in good agreement with the analytical results.  
We also find that the CCSS are in good agreement with the previously reported 
exciton binding energies from experimental and theoretical investigations 
(\autoref{fig:ccvss_compare}).
\begin{table*}[ht]
  \begin{center}
   \caption{\textbf{Exciton binding energies [meV] for CdSe quantum dots ranging in diameters from 
                    1.24nm to 20nm in size. The standard deviation $\sigma$ is reported in the last column.}}
   \label{tab:cdse_BE}
    \begin{tabular}{ l c c}
     \hline
        CdSe QD        & CCSS                  & $\sigma$\\ 
        Diameter [nm]  & Binding Energy [meV]  & [meV]     \\
        \hline
        %%0.50  & 2120.49  \\
        %%0.75  & 1415.39  \\
        %%1.00  & 1062.89  \\
        %1.24  &  854.86  \\
        %1.79  &  595.92  \\
        %2.76  &  388.00  \\
        %2.98  &  359.96  \\
        %3.28  &  327.46  \\
        %3.79  &  284.19  \\
        %4.80  &  224.89  \\
        %6.60  &  165.52  \\
        %10.0  &  110.03  \\
        %15.0  &   75.23  \\
        %20.0  &   57.44  \\
        %0.50  & 2120.49  \\
        %0.75  & 1415.39  \\
        %1.00  & 1062.89  \\
        1.24  &  855    & 1.24E-03 \\
        1.79  &  596    & 2.89E-03 \\
        2.76  &  388    & 8.24E-03 \\
        2.98  &  360    & 9.66E-03 \\
        3.28  &  327    & 1.22E-02 \\
        3.79  &  284    & 1.68E-02 \\
        4.80  &  225    & 3.19E-02 \\
        6.60  &  166    & 7.69E-02 \\
        10.0  &  110    & 2.72E-02 \\
        15.0  &   75.2  & 1.02E-02 \\
        20.0  &   57.4  & 2.64E-02 \\
      \hline
    \end{tabular}
  \end{center}
\end{table*}

\subsection{Excitation energy of  CdSe clusters using dynamic screening}
The developed CCSS method was applied for the calculation of 
excitation energy in small CdSe clusters.
The electronic excitation was described using 
electron-hole quasiparticle representation
and the electron-electron correlation effect was
incorporated using screened electron-hole interaction kernel. 
In this work, we have used the geminal screened electron-hole interaction 
kernel which has the following form,
\begin{align}
	K_\mathrm{eh}(1,2)
	=
	w(1,2)g(1,2) (1-P_{12}) ,
\end{align}
where $w(1,2)$ is residual electron-electron interaction operator,
$g(1,2)$ is explicitly-correlated Gaussian-type geminal operator, 
and the $P_{12}$ is the permutation operator
(\autoref{eq:w1}-\autoref{eq:p12} ), 
\begin{align}
\label{eq:w1}
	 \sum_{i<j} r_{ij}^{-1} - \sum_i v_\mathrm{HF}(i)
	 = \sum_{i<j} w(i,j)
\end{align}
\begin{align}
\label{eq:p12}
	P_{12} f(1,2) = f(2,1) .
\end{align}
Using diagrammatic perturbation theory, it can be shown that 
up to first-order in $g$, the excitation energy is given by the following
expression,\cite{bayne_2014_infinite}
\begin{align}
	\omega = \omega_0 + \langle i a \vert K_\mathrm{eh} \vert a i \rangle ,
\end{align}
where $\omega_0 $ is the independent quasiparticle excitation energy 
and is equal to the energy difference between the quasihole and quasielectron states $(\omega_0 = \epsilon_a-\epsilon_i)$.
The evaluation of the matrix element of $K_\mathrm{eh}$ was 
accomplished using the  developed CCSS method. 
The single-particle states were obtained
from Hartree-Fock calculations using LANL2DZ ECP basis. 
The Gaussian-type geminal function was expanded using three-term expansions
and the expansion coefficients are were obtained from
literature. The $b$ and $\gamma$ used in this work were 
0.867863 and 0.010425, respectively, for the binding energy calculation 
on the Cd$_{20}$Se$_{19}$ quantum dot. 
Excitation energy in the $\mathrm{Cd}_{20}\mathrm{Se}_{19}$
cluster using the CCSS method 
was calculated and was found to be $3.14 \pm 4 \times 10^{-4}$ eV.
This result was found to be in good agreement with the 
previously published excitation energy of 
$3.10$ eV obtained using pseudopotential+CI calculation. 
The application of the geminal-screened electron-hole interaction kernel 
method using analytical 
geminal AO integrals were computationally prohibitive 
for this system, however the developed CCSS method
allowed us to overcome the computational barrier (948 basis functions) and apply 
the explicity-correlated formulation to the calculation of excitation energy for this system. 
%--------------------------------------------------------------------------------------%
\section{Conclusion}
In conclusion, the development and implementation of the CCSS Monte Carlo method was presented. The CCSS method is a numerical integration scheme that uses Monte Carlo approach for calculation of MO integrals. The accuracy of Monte Carlo evaluation of integrals can be systematically improved by reducing the variance of the sample mean. In the CCSS method, we have combined both control-variate and stratified sampling strategies for variance reduction. The main feature of the CCSS method is that it avoids explicit AO-to-MO integral transformation for evaluation of the MO integrals. Consequently, it only requires value of the spatial MO at a given point which is readily obtained from the linear combination of the AOs. The use of stratified sampling in CCSS method is an important feature because the distribution of sampling points for each segment is optimized to minimize the overall variance. Computationally, this results in segments with higher variance are sampled proportionally more than segments with lower variance. Another feature of stratified sampling is that all instances of the calculated function are used for the estimation of the integral. This should be contrasted with rejection sampling Monte Carlo methods, where not all function evaluations contribute towards the estimation of the integral. This feature of stratified sampling has a direct impact on the efficiency of the overall calculation especially for cases where function evaluation is expensive. In the CCSS method, the variance of the sample mean was further reduced by introducing control-variate in the stratified sampling scheme. The control-variate in this approach plays an identical role as the importance function in Metropolis sampling. In this work, we have derived two different control-variates that are appropriate for MO integrals. The composite aspect of the CCSS method allows for evaluation of multiple MO integrals for the same stratified sampling step. Because the CCSS is a numerical method, it can be readily applied to complex kernels whose analytical integral in AO basis is not known. The developed CCSS method was applied for calculation of electron-hole matrix elements in the electron-hole explicitly correlated Hartree-Fock calculations and in the calculation of geminal-screened electron-hole interaction kernel. These methods were applied for investigation of excitonic properties of quantum dots. In both cases, the CCSS method not only allowed us to avoid the expensive AO-to-MO transformations but also allowed us to avoid calculation of AO integrals with R12 terms. 
\par
We believe that the CCSS method will be relevant for large-scale quantum mechanical calculations where AO-to-MO transformation is prohibitively expensive, calculations that are integral-direct where the AO integrals not pre-computed and stored, real-space and grid-based methods, many-body theories that use complex explicitly-correlated 2-electron, 3-electron, and higher n-electron operators for treating electron-electron correlation, and excited state calculations (such as CIS, Tamm-Dancoff, Bethe-Salpeter, GSIK and others) that require a small subset of MO integrals. 
%--------------------------------------------------------------------------------------%
\section{Acknowledgments}
We are grateful to National Science Foundation (CHE-1349892) and Syracuse University  for the financial support. 
%--------------------------------------------------------------------------------------%
\appendix
\section{Expectation value and variance}
\label{sec:appendixA}
We define a set of values $X$,
\begin{align}
	X = \{x_1, x_2, \dots, x_N \} .
\end{align}
The expectation value on set $X$ 
is defined by the following operation,
\begin{align}
	\mathbb{E}[X]
	&=
	\frac{1}{N} \sum_{i}^N x_i .
\end{align}
We also define the following common notations, 
\begin{align}
	aX    &\equiv   \{a x_1, a x_2, \dots, a x_N \} \\
	X+Y  &\equiv   \{ x_1+y_1,  x_2+y_2, \dots,  x_N+y_N \} \\
	XY  &\equiv   \{ x_1y_1,  x_2y_2, \dots,  x_N y_N \} .
\end{align}
Using this we can now write the following 
properties of $\mathbb{E}$,
\begin{align}
		\mathbb{E}[a X] &= a \mathbb{E}[X] \\
		\mathbb{E}[X+Y] &= \mathbb{E}[X] + \mathbb{E}[Y] .
\end{align}
These two properties can be combined into a single relationship,
\begin{align}
\label{eq:exp1}
	\mathbb{E}[\sum_\alpha^M a_\alpha X_\alpha]
	&=
	\sum_\alpha^M a_\alpha 
	\mathbb{E}[X_\alpha] .
\end{align}

The variance is defined as,
\begin{align}
	\mathbb{V}[X] = \mathbb{E}[X^2]  - \mathbb{E}[X]^2 .
\end{align}
Analogously, the covariance is defined as,
\begin{align}
	\mathbb{C}[X,Y] = \mathbb{E}[XY]  - \mathbb{E}[X]\mathbb{E}[Y] .
\end{align}
The variance has the following scaling property,
\begin{align}
	\mathbb{V}[aX] 
	=a^2 \mathbb{V}[X] .
\end{align}
\begin{proof}
	\begin{align}
	\mathbb{V}[aX] 
	&=
	\mathbb{E}[a^2X^2] - \mathbb{E}[aX]^2 \\
	&=
	a^2 \mathbb{E}[X^2] - a^2 \mathbb{E}[X]^2 \\
	&=
	a^2 \left(\mathbb{E}[X^2]  - \mathbb{E}[X]^2 \right) \\
	&=
	a^2 \mathbb{V}[X] 
\end{align}
\end{proof}
The variance of sum of  distributions is given by 
the following equation,
\begin{align}
	\mathbb{V}[\sum_\alpha^M a_\alpha X_\alpha] 
	=
	\sum_{\alpha \beta}^M
	a_\alpha a_\beta
	\mathbb{C}[X_\alpha,X_\beta] .
\end{align}
\begin{proof}
\begin{align}
\label{eq:var3}
	\mathbb{V}[\sum_\alpha^M a_\alpha X_\alpha] 
	&=
	\mathbb{E}[\sum_{\alpha \beta}^M a_\alpha a_\beta X_\alpha X_\beta]
	-\mathbb{E}[\sum_\alpha^M a_\alpha X_\alpha]^2 \\
	&=
	\sum_{\alpha \beta}^M
	a_\alpha a_\beta
	 \mathbb{E}[ X_\alpha X_\beta]
	 -	\sum_{\alpha \beta}^M
	a_\alpha a_\beta \mathbb{E}[X_\alpha] \mathbb{E}[X_\beta] \\
	&=
	\sum_{\alpha \beta}^M
	a_\alpha a_\beta
	 \mathbb{C}[ X_\alpha, X_\beta]	
\end{align}
\end{proof}
In case  $X_\alpha$ and  $X_\beta$
are uncorrelated then the covariance 
is zero,
\begin{align}
	\mathbb{C}[X_\alpha,X_\beta] 
	&= 0 
	\quad (\mathrm{for} \quad \alpha \neq \beta) .
\end{align}
The above expression reduces to,
\begin{align}
\label{eq:var4}
	\mathbb{V}[\sum_\alpha^M a_\alpha X_\alpha] 
	=
	\sum_{\alpha}^M
	a_\alpha^2 \mathbb{V}[X_\alpha]
	\quad (\textrm{for uncorrelated} \,\, X_\alpha ) .
\end{align}
The relationship between the variance in the sample mean and the 
variance of the underlying distribution can be obtained as follows,
\begin{align}
	\mathbb{V}[\mu]
	&=
	\mathbb{V}[\frac{1}{N} \sum_{i}^N X_i]
\end{align}
Because all the samples are uncorrelated,
\begin{align}
	\mathbb{V}[\mu]
	&=
	\frac{1}{N^2}
	\sum_{i}^N 
	\mathbb{V}[X_i]
\end{align}
Since $X_i$ is drawn for the same distributions,
all instances of $X_i$ have identical variance,
\begin{align}
\label{eq:var5}
	\mathbb{V}[\mu]
	&= 	\frac{1}{N^2}
	(N	\mathbb{V}[X]) \\
	&= 	
	\frac{\mathbb{V}[X]}{N}
\end{align}

%--------------------------------------------------------------------------------------%
\clearpage
\newpage
\bibliography{mgb_strat_samp}
%--------------------------------------------------------------------------------------%

\end{document}